\documentclass[preprint,aps]{revtex4}
\usepackage{amsmath}
\usepackage{amssymb}
\usepackage{color}

\begin{document}
{\flushright{TUHEP-TH-06152}}
\tighten \draft
\date{\today}
\color{black}
\title{Nonstandard Higgs in Electroweak Chiral Lagrangian }
\author{Li-Ming Wang, Qing Wang}

\address{Department of Physics,Tsinghua University,Beijing 100084,P.R.China}

\date{May 10, 2006}

\begin{abstract} We add a nonstandard higgs into the traditional
bosonic part of electroweak chiral Lagrangian, in purpose of
finding out the contribution to EWCL coefficients from processes
with internal line higgs particle. To construct the effective
Lagrangian with higgs, we use low energy expansion scheme and
write down all the independent terms conserving $SU(2)\times
U_Y(1)$ symmetry in the nonlinear representation which we show is
equivalent to the linear representation. Then we integrate out
higgs using loop expansion technique at 1-loop level,
contributions from all possible terms are obtained. We find up to
order of $p^4$ in low energy expansion, three terms,
$\mathcal{L}_5$, $\mathcal{L}_7$, $\mathcal{L}_{10}$ in EWCL are
important, for which the contributions from higgs can be further
expressed in terms of higgs partial decay width
$\Gamma_{h\rightarrow ZZ}$ and $\Gamma_{h\rightarrow WW}$. Higg
mass dependence of the coefficients in EWCL are discussed.

\bigskip
PACS number(s): 11.30.Rd, 12.39.Fe, 12.60.Fr, 14.80.Cp
\end{abstract}

\maketitle \color{black}
\section{Introduction}
The bosonic part of electroweak chiral Lagrangian (EWCL) first
introduced by A.Longhitano, T.Appelquist and Bernard in
Ref.~\cite{longhitano80,Appelquist80} is used to describe the impact
of higgs sector on the rest of the gauge theory, which supposes
higgs heavy and disappear from the electroweak interaction at low
energy scale $E\ll1Tev$. The most general form\cite{Appelquist93} of
this EWCL up to order of $p^4$ includes sixteen independent terms
which conserves $SU(2)\times U_Y(1)$ symmetry and contribute to
electroweak gauge boson self energy and
vertices\cite{Appelquist93,peskin92}. Because there are only
goldstones and gauge bosons in this effective theory, it provides an
economical phenomenological description of electroweak physics below
Tev energy scale.  The importantce of this higgsless description of
EW interaction is based on the fact that higgs has not been found in
present experiments below LEP bound $m_h>114.4 Gev$\cite{Eidelan04}
for which the situation will last until the higgs or other new
particles are found in future and the fact that the description of
higgs sector in SM is actually
problematic\cite{Dashen83,Susskind79}. Nowadays, in the situation
that LHC is going to run in 2007 and ILC is under active discussion,
EWCL plays more special role in particle physics, since even with
most optimistic estimation, discovery of new particles on LHC needs
at least three more years from now on. Before that time, the only
correct theory verified by experiments is this higgsless description
of EW interaction. In next few years, if we are not lucky in finding
new particles on LHC, EWCL may last even more time. During the time
that EWCL keep to be correct, the important question we need to
answer is: Can we test effects of new particles below their
thresholds? In the language of EWCL is to investigate the role of
new particles on the coefficients in the EWCL. Since the most
interesting new particle is higgs, in this paper, we focus our
attentions on investigating effects of single higgs in EWCL, the
effects from other possible new particles will be discussed
elsewhere. At present stage, due to lack of experiment data, it is
impossible to give these effects the quantitative estimations, but
we hope through our work, some qualitative features can be evaluated
out.

The possibility that an elementary higgs particle with mass higher
than 114.4 Gev always stands. Although we still know little about
the origin of the spontaneously electroweak symmetry breaking,
higgs mechanism tells that a heavy particle with nonzero vacuum
expectation value can always help us out of that confusion. Hence
it is interesting to consider the possibility that a nonstandard
higgs exists at $E>114.4$ Gev and to find out its possible
interactions with the rest of gauge theory. Once a effective
theory with higgs included is established above some energy scale,
 we can in principle integrate out all contributions from heavy
higgs to obtain an effective Lagrangian below that scale, i.e.,
the EWCL. The effects from contributions of the integrated out
heavy higgs resides in the coefficients of the EWCL, which reflect
effective interactions among goldstones and gauge bosons.

In literature, there are papers to discuss the issue we are
interested \cite{Dittmaier95,Gasser84,Gasser89,Donoghue89}. The
theory for higgs they started with are limited either in standard
model (SM) or some simplified higgs models which has no custodial
symmetry breaking. In this paper, we start from the most general
theory--EEWCL which is an extended EWCL theory with higgs included
in to make our investigations. This article is organized as follows:
in Sec.II, two different descriptions of the EW interaction, linear
and nonlinear representations, are discussed and the equivalence
between these two is demonstrated. In Sec.III, all possible
independent interaction terms among higgs, goldstones and gauge
bosons are introduced and a complete set EEWCL up to certain order
of the low energy expansion which will finally contribute to $p^4$
order EWCL at one loop level is constructed. After that, in Sec.IV,
we use loop expansion scheme\cite{Jackiw74} to integrate out the
heavy higgs to obtain the contributions from it to the traditional
EWCL. Finally, some integration results are listed and discussed.
\section{Equivalence between linear and non-linear representation}
Higgs is first introduced into electroweak sector by Weinberg and
Salam\cite{Weinberg67} as a SU(2) isospin doublet,
\begin{eqnarray}
\Phi=\frac{1}{\sqrt{2}}\left(
                         \begin{array}{c}
                           \phi^+ \\
                           \phi^0 \\
                         \end{array}
                       \right)\;,
\end{eqnarray}
which transforms linearly under $SU(2)$ and $U_Y(1)$ gauge group
action. And all the interactions related to higgs doublet can be
constructed readily with simple lie algebras. With this linear
form of higgs doublet, the bosonic part of higgs effective
Lagrangian can be written as
\begin{eqnarray}
\mathcal{L}_{eff}=\mathcal{L}_{SM}+\sum\limits_n\frac{f_n}{\Lambda_H^2}\mathcal{O}_n\;.
\end{eqnarray}
The SM Lagrangian contains terms up to dimension four operators,
thus nonstandard higgs interaction starts from dimension six. In
dimension six part of Lagrangian,  12 independent operators form a
basis set. In the notation of
Ref.~\cite{Buchmuller86,hagiwara96,Gonzalez-Garcia99}, they are
\begin{eqnarray}
\mathcal{O}_{DW}&=&Tr([D_{\mu},\hat W_{\nu\rho}][D^{\mu},\hat
W^{\nu\rho}])\nonumber\\
\mathcal{O}_{DB}&=&-\frac{g'^2}{2}\partial_{\mu}B_{\nu\rho}\partial^{\mu}B^{\nu\rho}\nonumber\\
\mathcal{O}_{BW}&=&\Phi^+\hat B_{\mu\nu}\hat W^{\mu\nu}\Phi\nonumber\\
\mathcal{O}_{\Phi,1}&=&\left[(D_{\mu}\Phi)^+\Phi\right]\left[\Phi^+D^{\mu}\Phi\right]\nonumber\\
\mathcal{O}_{WWW}&=&Tr(\hat W_{\mu\nu}\hat W^{\nu\rho}\hat
W_{\rho}^{\mu})\nonumber\\
\mathcal{O}_{WW}&=&\Phi^+\hat W_{\mu\nu}\hat W^{\mu\nu}\Phi\nonumber\\
\mathcal{O}_{BB}&=&\Phi^+\hat B_{\mu\nu}\hat B^{\mu\nu}\Phi\nonumber\\
\mathcal{O}_{W}&=&(D_{\mu}\Phi)^+\hat W^{\mu\nu}(D_{\nu}\Phi)\nonumber\\
\mathcal{O}_{B}&=&(D_{\mu}\Phi)^+\hat B^{\mu\nu}(D_{\nu}\Phi)\nonumber\\
\mathcal{O}_{\Phi,2}&=&\frac{1}{2}\partial_{\mu}(\Phi^+\Phi)\partial^{\mu}(\Phi^+\Phi)\nonumber\\
\mathcal{O}_{\Phi,3}&=&\frac{1}{3}(\Phi^+\Phi)^3\nonumber\\
\mathcal{O}_{\Phi,4}&=&(\Phi^+\Phi)\left[(D_{\mu}\Phi)^+(D^{\mu}\Phi)\right]\;.\label{Linear}
\end{eqnarray}
The covariant derivative D is given by
\begin{eqnarray}
D_{\mu}=\partial_{\mu}+igT^aW^a_{\mu}+ig'YB_{\mu}\;,
\end{eqnarray}
where $g$ is SU(2) coupling with
$Tr(T^aT^b)=\frac{1}{2}\delta^{ab}$, $g'$ is $U_Y(1)$ coupling and Y
is the hypercharge operator. Define
\begin{eqnarray}
\hat W_{\mu\nu}=igT^aW^a_{\mu\nu}\hspace{2cm}\hat
B_{\mu\nu}=ig'B_{\mu\nu}\;,
\end{eqnarray}
hence
\begin{eqnarray}
[D_{\mu},D_{\nu}]=\hat W_{\mu\nu}+\hat B_{\mu\nu}\;.
\end{eqnarray}
Using these twelve operators, effective interactions between higgs
doublet field and gauge bosons could be calculated directly.
Operators in dimension 8 are given in Ref.~\cite{ARZT95}.

Different from the doublet representation, EW chiral
Lagrangian(EWCL)\cite{longhitano80,Appelquist80,Appelquist93} takes
a non-linear representation, with goldstone field given by,
\begin{eqnarray}
U=e^{i\frac{\tau^i\pi^i}{2f}},\hspace{2cm}i=1,2,3
\end{eqnarray}
where $\tau^i,i=1,2,3$ are three pauli matrices. In the notation
of Ref.[2], there are two independent terms in $p^2$ order,
\begin{eqnarray}
\mathcal{L}^{(2)}=-\frac{f^2}{4}Tr[(D_{\mu}U)^+(D^{\mu}U)]+\frac{\beta_1f^2}{4}[Tr(TV_{\mu})]^2\;,\label{L2}
\end{eqnarray}
where
\begin{eqnarray}
D_{\mu}U=\partial_{\mu}U+ig\frac{\tau^a}{2}W^a_{\mu}U-ig'U\frac{\tau^3}{2}B_{\mu}\;.
\end{eqnarray}
T operator in the second terms of (\ref{L2}) breaks custodial
symmetry into $SU(2)\times U_Y(1)$ even in the absence of $U_Y(1)$
gauge coupling, with
\begin{eqnarray}
&&V_{\mu}=(D_{\mu}U)U^+\hspace{1cm}T=U\tau^3U^+\nonumber\\
&&W_{\mu\nu}=\partial_{\mu}W_{\nu}-\partial_{\nu}W_{\mu}+ig[W_{\mu},W_{\nu}]\nonumber\\
&&B_{\mu\nu}=\partial_{\mu}B_{\nu}-\partial_{\nu}B_{\mu}\label{buildingblock}\;.
\end{eqnarray}
All these four building operators in (\ref{buildingblock}) are $SU(2)$ covariant and $U_Y(1)$ invariant. On this
basis, fourteen independent terms in $p^4$ order are given by,
\begin{eqnarray}
\mathcal{L}_1&\equiv&\frac{\alpha_1}{2}gg'B_{\mu\nu}Tr(TW^{\mu\nu})=\frac{\alpha_1}{2}gg'l_4^1\nonumber\\
\mathcal{L}_2&\equiv&\frac{i\alpha_2}{2}g'B_{\mu\nu}Tr(T[V^{\mu},V^{\nu}])=\frac{i\alpha_2}{2}g'l_4^2\nonumber\\
\mathcal{L}_3&\equiv&i\alpha_3gTr(W_{\mu\nu}[V^{\mu},V^{\nu}])=i\alpha_3gl_4^3\nonumber\\
\mathcal{L}_4&\equiv&\alpha_4[Tr(V_{\mu}V_{\nu})]^2=\alpha_4l_4^4\nonumber\\
\mathcal{L}_5&\equiv&\alpha_5[Tr(V_{\mu}V^{\mu})]^2=\alpha_5l_4^5\nonumber\\
\mathcal{L}_6&\equiv&\alpha_6Tr(V_{\mu}V_{\nu})Tr(TV^{\mu})Tr(TV^{\nu})=\alpha_6l_4^6\nonumber\\
\mathcal{L}_7&\equiv&\alpha_7Tr(V_{\mu}V^{\mu})Tr(TV_{\nu})Tr(TV^{\nu})=\alpha_7l_4^7\nonumber\\
\mathcal{L}_8&\equiv&\frac{1}{4}\alpha_8g^2[Tr(TW_{\mu\nu})]^2=\frac{1}{4}\alpha_8g^2l_4^8\nonumber\\
\mathcal{L}_9&\equiv&\frac{i}{2}\alpha_9gTr(TW_{\mu\nu})Tr(T[V^{\mu},V^{\nu}])=\frac{i}{2}\alpha_9gl_4^9\nonumber\\
\mathcal{L}_{10}&\equiv&\frac{1}{2}\alpha_{10}[Tr(TV_{\mu})Tr(TV_{\nu})]^2=\frac{1}{2}\alpha_{10}l_4^{10}\nonumber\\
\mathcal{L}_{11}&\equiv&\frac{1}{2}\alpha_{11}g\epsilon_{\mu\nu\rho\lambda}Tr(TV^{\mu})Tr(V^{\nu}W_{\rho\lambda})=\frac{1}{2}\alpha_{11}gl_4^{11}\nonumber\\
\mathcal{L}_{12}&\equiv&\alpha_{12}gTr(TV^{\mu})Tr(V_{\nu}W^{\mu\nu})=\alpha_{12}gl_4^{12}\nonumber\\
\mathcal{L}_{13}&\equiv&\alpha_{13}gg'\epsilon_{\mu\nu\rho\lambda}B^{\mu\nu}Tr(TW^{\rho\lambda})=\alpha_{13}gg'l_4^{13}\nonumber\\
\mathcal{L}_{14}&\equiv&\alpha_{14}g^2\epsilon_{\mu\nu\rho\lambda}Tr(TW^{\mu\nu})Tr(TW^{\rho\lambda})=\alpha_{14}g^2l_4^{14}\label{Nonlinear}\;.
\end{eqnarray}
$\mathcal{L}_{12}$, $\mathcal{L}_{13}$ and $\mathcal{L}_{14}$ are
three CP violation terms.

 Actually, there is a natural connection
between the doublet linear representation given in (\ref{Linear})
and Non-linear
representation given in (\ref{Nonlinear}), which we would show explicitly below.\\
In unitary gauge, higgs doublet is parameterized as
\begin{eqnarray}
\Phi=\frac{1}{\sqrt{2}}e^{i\pi^a\tau^a}\left(
                         \begin{array}{c}
                           0 \\
                           h+v \\
                         \end{array}
                       \right)\;.
\end{eqnarray}
Define the charge conjugation of $\Phi$
\begin{eqnarray}
\Phi^c\equiv i\tau^2\Phi^*=    \left(\begin{array}{c}
                           h+v \\
                           0 \\
                         \end{array}
                       \right)\;.
\end{eqnarray}
Set
\begin{eqnarray}
\Sigma\equiv \left(
               \begin{array}{cc}
                 \Phi^c & \Phi \\
               \end{array}
             \right)=\frac{h+v}{\sqrt{2}}e^{i\pi^a\tau^a}\equiv
             \frac{h+v}{\sqrt{2}}U\;,
\end{eqnarray}
where U is defined as
\begin{eqnarray}
U\equiv e^{i\pi^a\tau^a}\;.
\end{eqnarray}
By doing some SU(2) algebras, we find following connections
between the two representations
\begin{eqnarray}
2(D_{\mu}\Phi)^+\Phi&=&\partial_{\mu}h^2+h^2Tr(TV_{\mu})\nonumber\\
2\Phi^+W_{\mu\nu}\Phi&=&h^2Tr(TW_{\mu\nu})\nonumber\\
2(D_{\mu}\Phi)^+(D_{\nu}\Phi)&=&h^2[Tr(TV_{\mu}V_{\nu})-Tr(V_{\mu}V_{\nu})]+2(\partial_{\mu}h)(\partial_{\nu}h)\nonumber\\
2(D_{\mu}\Phi)^+W^{\mu\nu}(D_{\nu}\Phi)&=&h^2Tr(W^{\mu\nu}V_{\mu}V_{\nu})-(\partial_{\mu}h^2)Tr(W^{\mu\nu}V_{\nu})\nonumber\\
2\Phi^+W^{\nu\rho}(D^{\mu}\Phi)&=&h^2[Tr(TV^{\mu}W^{\nu\rho})+Tr(V^{\mu}W^{\nu\rho})]\nonumber\\
2(D^{\mu}\Phi)^+W^{\nu\rho}\Phi&=&h^2[Tr(TV^{\mu}W^{\nu\rho})-Tr(V^{\mu}W^{\nu\rho})]\label{Equivalence}\;.
\end{eqnarray}
Here higgs field h and goldstone field U are defined as
\begin{eqnarray}
h^2\equiv\det\Sigma\hspace{1cm} \Sigma\equiv hU\;.
\end{eqnarray}

Thus higgs here is a $SU(2)\times U(1)$ scalar, denoting the module
freedom of higgs doublet, while goldstone U denotes the rotation
angle of EW gauge transformation, which is similar to the case in
the chiral Lagrangian of strong interaction, where a scalar meson
$\sigma$ denotes the module and eight goldstone U denotes angle of
the strong chiral transformation\cite{wang00}. Note that $h$ is not
exactly the parameter $'h'$ of higgs doublet in the unitary gauge,
but differs in a factor of $\sqrt{2}$.

We can also express all possible interacting terms in the
non-linear representation with higgs doublet as followings:
\begin{eqnarray}
&&Tr(TV_{\mu})=(\Phi^+\Phi)^{-1}[2(D_{\mu}\Phi)^+\Phi-\partial_{\mu}(\Phi^+\Phi)]\nonumber\\
&&Tr(TW_{\mu\nu})=2(\Phi^+\Phi)^{-1}[\Phi^+W_{\mu\nu}\Phi]\nonumber\\
&&Tr(V_{\mu}V_{\nu})=\frac{1}{2}(\Phi^+\Phi)^{-2}\partial_{\mu}(\Phi^+\Phi)\partial_{\nu}(\Phi^+\Phi)-(\Phi^+\Phi)^{-1}[(D_{\mu}\Phi)^+(D_{\nu}\Phi)+h.c.]\nonumber\\
&&Tr(TV_{\mu}V_{\nu})=(\Phi^+\Phi)^{-1}[(D_{\mu}\Phi)^+(D_{\nu}\Phi)-h.c.]\nonumber\\
&&Tr(V^{\mu}W^{\nu\rho})=(\Phi^+\Phi)^{-1}[-(D^{\mu}\Phi)^+W^{\nu\rho}\Phi+h.c.]\nonumber\\
&&Tr(TV^{\mu}W^{\nu\rho})=(\Phi^+\Phi)^{-1}[(D^{\mu}\Phi)^+W^{\nu\rho}\Phi+h.c.]\nonumber\\
&&Tr(W^{\mu\nu}V_{\mu}V_{\nu})=2(\Phi^+\Phi)^{-1}[(D_{\mu}\Phi)^+W^{\mu\nu}(D_{\nu}\Phi)]+(\Phi^+\Phi)^{-2}\partial_{\mu}(\Phi^+\Phi)[-(D^{\mu}\Phi)^+W^{\nu\rho}\Phi+h.c.]\nonumber\\
\end{eqnarray}
In the path integral system, the change of variables induces a
determinant factor to the generating functional $\mathcal{Z}$,
\begin{eqnarray}
\mathcal{Z}&=&\int\mathcal{D}W_{\mu}\mathcal{D}B_{\mu}\mathcal{D}U\mathcal{D}h
\exp\{iS'[W_{\mu}^a,U,h]\}\det\{i\delta^{(4)}(0)(h+v)\}
\end{eqnarray}
The determinant can be written in the exponential form.
Correspondingly, the lagrangian density transforms as
\begin{eqnarray}
\mathcal{L}\rightarrow\mathcal{L}'+\delta^{(4)}(0)\ln\,(h+v)
\end{eqnarray}
This determinant which contains quartic divergences is necessary to
cancel exactly the quartic divergences brought into by the
longitudinal part of gauge boson.\cite{Woodhouse73,Grosse-Knetter93}
However in our discussion the logarithm term is totally absorbed in
the free parameters of the higgs potential.\\
According to above facts and algebra relations, any terms written in
a linear representation with doublet $\Phi$ can be transformed into
terms written in a nonlinear one with scalar $h$ and $U$, and vice
verse. In this sense, an effective theory written in two different
representations are actually the same. For example, bosonic sector
of SM could be transformed as
\begin{eqnarray}
\mathcal{L}_{SM}=\frac{1}{2}(\partial_{\mu}h)^2-\frac{(h+v)^2}{4}Tr(V_{\mu}V^{\mu})+\mu^2(h+v)^2-\lambda
(h+v)^4\label{SM}\;.
\end{eqnarray}

\section{An extended EW chiral Lagrangian
(EEWCL)} From relations in (\ref{Equivalence}), we know higgs
could be added into traditional EW chiral Lagrangian in a simple
way. According to the principle of phenomenological Lagrangian by
Weinberg in Ref.~\cite{Weinberg79}, We just write down all
possible independent terms conserving $SU(2)\times U(1)$ symmetry,
and arrange them order by order. The construction of the extended
EW chiral Lagrangian can be done in quite a straight way, once we
know how to deal with higgs in power counting. Traditional EWCL
takes Weinberg's power counting scheme\cite{Weinberg79}, in which
goldstone U is counted as $p^0$ order, while $\partial_{\mu}$ and
gauge boson fields as $p^1$ order. But here we have no good reason
to designate higgs any certain power, in that we suppose higgs
interacts with gauge bosons in arbitrary way permitted by
$SU(2)\times U(1)$ symmetry. Because our purpose is to find out
the effective contribution from higgs to EWCL couplings, thus a
better way is to take a expansive view on this heavy higgs field,
i.e.
\begin{eqnarray}
h=h^{(0)}+h^{(2)}+h^{(4)}+\ldots
\end{eqnarray}
However, at present we needn't care much about this power counting
problem, as long as all terms which would contribute to EWCL
couplings are included. Through complicated calculation, following
terms are found out to be complete and independent,
\begin{eqnarray}
\mathcal{L}&=&\mathcal{L}^{(0)}+\mathcal{L}^{(2)}+\mathcal{L}^{(4)}\label{EEWCL0}\nonumber\\
\mathcal{L}^{(at~least~0)}&=&-V(h)\nonumber\\
\mathcal{L}^{(at~least~2)}&=&\frac{1}{2}(\partial_{\mu}h)^2+C_1(h)A_{\mu}^2+C_2(h)tr(V_{\mu}^2)\nonumber\\
\mathcal{L}^{(at~least~4)}&=&C_3^i(h)l^i_4+C_4^j(h)(\partial_{\mu}h)l_3^{j\mu}+C_5^k(h)(\partial_{\mu}h)(\partial_{\nu}h)l_2^{k\mu\nu}\nonumber\\
&&+C_6(h)(\partial_{\mu}h)^2(\partial_{\nu}h)A^{\nu}+C_7(h)(\partial_{\mu}h)^4\label{Lagrangian}\nonumber\\
\mathcal{L}^{(at~least
~6)}&=&C_8^l(h)(\partial_{\mu}h)(\partial_{\nu}h)l^{l\mu\nu}_4\;,
\end{eqnarray}
in which
\begin{eqnarray}
A_{\mu}&=&tr(TV_{\mu})\;,
\end{eqnarray}
$V(h)$ is some arbitrary potential of $h$, $C_n(h)$ are
coefficient functions depending on higgs field. $l^{i}_4,
i=1,2,\ldots,14$, are $p^4$ order operators defined in
(\ref{Nonlinear}), $l_3^{j\mu},j=1,2,\ldots,7$ and
$l_2^{k\mu\nu},k=1,2$, are $p^3$ and $p^2$ order tensors depending
on goldstone U, gauge field W and B, they are given by
\begin{eqnarray}
&&l^{1\mu\nu}_2=Tr(TV^{\mu})Tr(TV^{\nu})\hspace{1cm}
l^{2\mu\nu}_2=Tr(V^{\mu}V^{\nu})\nonumber\\
&&l^{1\mu}_3=Tr(TV^{\mu})Tr(V^{\nu}V_{\nu})\hspace{1cm}
l^{2\mu}_3=Tr(TV^{\nu})Tr(V^{\mu}V_{\nu})\nonumber\\
&&l^{3\mu}_3=Tr(TV^{\nu})Tr(TV^{\mu}V_{\nu})\hspace{0.7cm}
l^{4\mu}_3=Tr(TV_{\nu})Tr(TW^{\mu\nu})\nonumber\\
&&l^{5\mu}_3=B^{\mu\nu}Tr(TV_{\nu})\hspace{2cm}
l^{6\mu}_3=Tr(TW^{\mu\nu}V_{\nu})\nonumber\hspace{0.5cm}
l^{7\mu}_3=Tr(W^{\mu\nu}V_{\nu})\;.
\end{eqnarray}
The last term $l^{l\mu\nu}_4$ includes all possible forms of tensors with symmetric $\mu\nu$ indices in $p^4$
order.
\begin{eqnarray}
l^{1\mu\nu}_4&=&B^{\mu}_{\rho}Tr(TW^{\nu\rho})\nonumber\\
l^{2\mu\nu}_4&=&B^{\mu}_{\rho}Tr(T[V^{\nu},V^{\rho}])\nonumber\\
l^{3\mu\nu}_4&=&Tr(W^{\mu}_{\rho}[V^{\nu},V^{\rho}])\nonumber\\
l^{4\mu\nu}_4&=&Tr(V^{\mu}V_{\rho})Tr(V^{\nu}V^{\rho})\nonumber\\
l^{5\mu\nu}_4&=&Tr(V^{\mu}V^{\nu})Tr(V^{\rho}V_{\rho})\nonumber\\
l^{6\mu\nu}_4&=&Tr(V^{\mu}V^{\nu})Tr(TV^{\rho})Tr(TV_{\rho})\nonumber\\
l^{7\mu\nu}_4&=&Tr(V^{\mu}V_{\rho})Tr(TV^{\nu})Tr(TV^{\rho})\nonumber\\
l^{8\mu\nu}_4&=&Tr(TW^{\mu}_{\rho})Tr(TW^{\mu\rho})\nonumber\\
l^{9\mu\nu}_4&=&Tr(TW^{\mu}_{\rho})Tr(T[V^{\mu},V^{\rho}])\nonumber\\
l^{10\mu\nu}_4&=&Tr(TV^{\mu})Tr(TV^{\nu})Tr(TV^{\rho})Tr(TV_{\rho})\nonumber\\
l^{12\mu\nu}_4&=&Tr(TV^{\mu})Tr(V_{\rho}W^{\nu\rho})\;.
\end{eqnarray}
Two terms $C_9(h)Tr(W_{\mu\nu}W^{\mu\nu})$ and
 $C_{10}(h)B_{\mu\nu}B^{\mu\nu}$ are not included in
(\ref{EEWCL0}) since their contributions could be represented by
redefinition of gauge boson field and the gauge
coupling\cite{hagiwara96}.

 In (\ref{EEWCL0}), besides the old
terms $l_4^{11}$,$l_4^{12}$,$l_4^{13}$, there are other new terms
violating CP, whose couplings are $C_6$, $C_4^3$, $C_4^4$, $C_4^6$
and $C_8^{12}$, respectively. It is interesting that these CP
violating terms might contribute to CP conserving effective
couplings in EWCL.
\section{Integrate out higgs}

Suppose higgs is heavy, we want to integrate out higgs in this
EEWCL to obtain the contributions to the effective couplings of
the rest part of EW interaction. Because we stick to the
equivalence between the linear and non-linear representation, we
need a nonzero vacuum condensation of higgs field to induce mass
to gauge bosons.  When all external fields $V_{\mu}$, $W_{\mu}$
and $B_{\mu}$ vanish, we are left with a theory of higgs
self-interaction. The vacuum expectation value(vev) is determined
by the stationery equation of higgs potential
\begin{eqnarray}
0=\delta V(h)\Rightarrow V'(h)\bigg|_{h=v}=0\label{vev}\;.
\end{eqnarray}
If we first complete the renormalization of this higgs potential,
then find the solution of the stationery equation of the
renormalized potential, we get the physical vev of higgs which
includes contributions from all quantum corrections. We take v here
a free parameter because higgs potential itself is totally free. And
the physical higgs $\tilde h$ with zero vev is given by
\begin{eqnarray}
h=\tilde h+(v+\delta v)\;,
\end{eqnarray}
here $\delta v$ denotes loop correction to vev, which is
determined by
\begin{eqnarray}
\frac{d}{dh}[V_{tree}(h)+V_{loop}(h)]\bigg|_{h=v+\delta
v}=0\;,\label{vevloop}
\end{eqnarray}
Suppose higgs potential $V(h)$ is expanded as
\begin{eqnarray}
V_{tree}(h)=V_{tree}(v)+\frac{1}{2}m^2(h-v)^2+\frac{1}{6}am
(h-v)^3+\frac{1}{12}b (h-v)^4+\ldots\;,\label{Vtree}
\end{eqnarray}
and
\begin{eqnarray}
V_{loop}(h)=V_{loop}(v)+\delta c\cdot m^3(h\!-\!v)+\frac{1}{2}\delta
m^2(h\!-\!v)^2+\frac{1}{6}\delta am (h\!-\!v)^3+\frac{1}{12}\delta b
(h\!-\!v)^4+\cdots~~~~\;,\label{Vloop}
\end{eqnarray}
thus, with (\ref{Vtree}), (\ref{Vloop}) and (\ref{vevloop}), we are
left with following equation of the loop correction for vev
\begin{eqnarray}
\frac{1}{3}(b+\delta b)y^3+\frac{1}{2}(a+\delta
a)y^2+(1+\frac{\delta m^2}{m^2})y+\delta c=0\;,\label{cubic}
\end{eqnarray}
where $y\equiv\delta v/m$, to simplify the solution of this cubic
equation, we introduce some expansion parameter which will
decompose the solution order by order; one reasonable assumption
is to impose some small '$\lambda$' dependence on these  the
couplings in eq.(\ref{cubic}), which is the characteristic
coupling strength in SM higgs potential.
\begin{eqnarray}
 a\sim \lambda^{1/2}\quad,b\sim \lambda^1,\quad m^2\sim
 \lambda^0\label{lambdatree}
\end{eqnarray}
and on the one loop corrections
\begin{eqnarray}
\delta a\sim \lambda^{3/2}\quad\delta b\sim \lambda^2, \quad\delta
m^2\sim \lambda^1 \quad\delta c\sim
\lambda^{1/2}\;.\label{lambdaloop}
\end{eqnarray}
We comment that this kind of $\lambda$ dependence is naturally
supported by our power counting rule, i.e., the assumption of
smaller coefficients of higher order operators in the low energy
expansion because $m^2$, $am$ and $b$ are coefficients of $h^2$,
$h^3$ and $h^4$, respectively. In next section (\ref{A}) we will see
these three terms belong to $p^4$, $p^6$ and $p^8$ orders in our
higgs power counting system. With (\ref{lambdatree}) and
(\ref{lambdaloop}), the leading order of $\lambda$ dependence of
$\delta v$ is found to be $\lambda^{1/2}$, which is again a natural
result. Substitute them into (\ref{cubic}), we find
\begin{eqnarray}
\frac{1}{3}\delta
by^3\lambda^{7/2}+\frac{1}{3}by^3\lambda^{5/2}+\frac{1}{2}\delta
ay^2\lambda^2+(\frac{1}{2}ay^2+\frac{\delta
m^2}{m^2}y)\lambda^{3/2}+(y+\delta
c)\lambda^{1/2}=0\;.\label{lambdaorder}
\end{eqnarray}
The leading order of $\lambda$ of this equation is $1/2$, thus,
\begin{eqnarray}
y=-\delta c\;.
\end{eqnarray}
Detail one loop calculation gives
\begin{eqnarray} \delta
c=\frac{a}{32\pi^2}(\frac{1}{\epsilon}-\gamma+1+\ln\frac{4\pi\mu^2}{m^2})\hspace{1cm}
\delta v=-\frac{a\,
m}{32\pi^2}(\frac{1}{\epsilon}-\gamma+1+\ln\frac{4\pi\mu^2}{m^2})\;.
\end{eqnarray}
Beyond leading order of $\lambda$ counting, one interesting case
is $v|_{tree}=0$, $a=0$ and $m=0$, which represent the situation
that electroweak symmetry donot violates at tree level of EEWCL.
 E.q.(\ref{cubic}) in this situation becomes
\begin{eqnarray}
\frac{1}{3}(b+\delta b)y^3+\frac{\delta m^2}{m^2}y=0\;.
\end{eqnarray}
Beside the trivial solution of $y=0$, the nonzero leading dependence
of $\lambda$ is $y\sim \lambda^0$,
\begin{eqnarray}
y(\frac{1}{3}by^2+\frac{\delta
m^2}{m^2})\lambda^1+\frac{1}{3}\delta by^3\lambda^2=0\;,
\end{eqnarray}
which gives
\begin{eqnarray}
(\delta v)^2=-\frac{3\delta m^2}{b}\label{radicorr}\;.
\end{eqnarray}
This is actually the theory of massless quadratically
self-interacting meson field by S.Coleman and E.Weinberg in
Ref.~\cite{Coleman73}. It is quite straightforward using the
following calculation result (\ref{lnD}) to check (\ref{radicorr})
agrees with the result in Ref.~\cite{Coleman73}. Since $\delta
m^2\propto b$, (\ref{radicorr}) tells that this one loop vev
radiative correction is independent of the self-interacting
coupling as long as it is small enough to allow the perturbation.
However, this massless mode is not included in following of our
discussion because mass term in general case is more important
than the self-interacting parts of higgs potential according to
our higgs power counting.
\subsection{higgs power counting}\label{A}
Before going any further, we turn back to the question raised in
last section about the power counting problem of higgs field. It's
condensation v is disconnected to any external source with explicit
order, hence in order to include into theory all information from v,
it should be counted as order of $p^0$ . The physical part of higgs
interacts with external sources, hence it should be counted as at
least $p^2$ order. That is to say,
\begin{eqnarray}
h^{(0)}=v+\delta v,\hspace{1cm}\tilde h=h^{(2)}+h^{(4)}+\ldots\;.
\end{eqnarray}
This is consistent with the power counting in
Ref.~\cite{Gasser84,Gasser85}, where the scalar source is also
counted as $p^2$ order. On the other hand, because we care
contributions from heavy higgs up to $p^4$ order, to which the
only way of $h^{(4)}$ to contribute is in a linear form, which
vanishes due to the vacuum condensation condition (the
contribution to $p^2$ EWCL coupling from $h^{(2)}$
 vanishes due to the same reason).

From now on we use h to denote $\tilde h$, define $x\equiv h/m$.
The Lagrangian (\ref{EEWCL0}) is reparameterized  as
\begin{eqnarray}
\mathcal{L}_{EEWCL}&=&\frac{1}{2}(\partial_{\mu}h)^2-\frac{1}{2}m^2(1-a\delta c)h^2-m^4\tilde{V}(x)\nonumber\\
&&+m^2F_1(x)A_{\mu}^2+m^2F_2(x)Tr(V_{\mu}^2)\nonumber\\
&&+G_0^i(x)l_4^i+G_1^j(x)(\partial_{\mu}x)l_3^{j\mu}+\frac{1}{2}G_2^k(x)(\partial_{\mu}x)(\partial_{\nu}x)l_2^{k\mu\nu}\nonumber\\
&&+G_3(x)(\partial_{\mu}x)^2(\partial_{\nu}x)A^{\nu}+G_4(x)(\partial_{\mu}x)^4
+\frac{1}{2}m^{-2}G_5^l(x)(\partial_{\mu}x)(\partial_{\nu}x)l_4^{l\mu\nu}\label{EEWCL}\;.
\end{eqnarray}
Where
\begin{eqnarray}
\tilde{V}(x)&=&\frac{1}{6}amx^3+\frac{1}{12}bx^4+\ldots\nonumber\\
F_1(x)&=&(f_1-f_3\delta c)+(f_3-f_5\delta c)x+f_5x^2+\dots\nonumber\\
F_2(x)&=&(f_2-f_4\delta c)+(f_4-f_6\delta c)x+f_6x^2+\dots\nonumber\\
G_{\alpha}(x)&=&(g_{\alpha}-g'_{\alpha}\delta
c)+(g'_{\alpha}-g''_{\alpha}\delta
c)x+\frac{1}{2}(g''_{\alpha}-g'''_{\alpha}\delta c)x^2+\ldots,
\hspace{0.5cm}\alpha=0,1,\ldots,5
\end{eqnarray}
$f_i,i=1,\ldots,6$ and $g_{\alpha},\alpha=0,1,\ldots,5$ are
functions of v, all of them are free parameters because v is free,
and terms containing $\delta c$ denote loop correction from vev.

After the renormalization procedure is performed, all of the
quantum corrections from higher order operators are included, we
have such following equation of motion given by $p^4$ Lagrangian,
\begin{eqnarray}
0=\delta \mathcal{L}^{(4)}\Rightarrow\alpha h^{(2)}_c+\beta
A_{\mu}^2+\gamma Tr(V_{\mu}^2)=0\;,
\end{eqnarray}
where $\alpha$, $\beta$, $\gamma$ are renormalized parameters,
with which we get the classic solution of higgs. Also, we have a
similar equation from some higher $p^n$ order,
\begin{eqnarray} 0=\delta
\mathcal{L}^{(n+2)}\Rightarrow\partial_{\mu}\partial^{\mu}h^{(n-2)}\sim
h^{(n)}+\mathrm{other~ p^n~terms}\label{MQH}\;.
\end{eqnarray}
On the other hand, note that $p^2$ Lagrangian is disconnected to
higgs, it gives a $p^2$ equation of motion for
$V_{\mu}$\cite{Appelquist93}.
\begin{eqnarray}
D_{\mu}V^{\mu}=A\left[\partial_{\mu}Tr(TV^{\mu})\right]T+BTr(TV_{\mu})[V^{\mu},T]+O(p^4)\;,
\end{eqnarray}
$A$, $B$ are some coefficients, from this equation of motion, one
obtain
\begin{eqnarray}
\left[\partial_{\mu}Tr(TV^{\mu})\right]=Tr(D_{\mu}(TV^{\mu}))=Tr([V_{\mu},T]V^{\mu})+Tr(TD_{\mu}V^{\mu})=2A\left[\partial_{\mu}Tr(V^{\mu})\right]~~~\;,
\end{eqnarray}
which leads to
\begin{eqnarray}
\partial_{\mu}A^{\mu}=O(p^4)\label{MQU}\;.
\end{eqnarray}
(\ref{MQH}) and (\ref{MQU}) are two  equations of motion we use to
simplify higher order operators.

 The difference between this higgs power counting rule and Weinberg's power counting scheme
lies in the fact that operators in higher order would contribute
to lower orders through loop correction. This might indicate that
higgs loop contribution to EWCL should be rather small compared
with its tree level values. Note that 1-loop contribution from
$G_3$ in the low energy expansion vanishes using (\ref{MQU}). Up
to $p^4$ order, One loop contribution from $G_1$ coupling is a
total differential term. One loop contribution from $G_5$ could be
totally represented by $G_0$. There's no 1-loop contribution from
$G_4$. We arrange all independent terms order by orders as follows
which could contribute to $p^2$ or $p^4$ order EWCL coefficients,
\begin{eqnarray}
\mathcal{L}^{(2)}&=&m^2\left[(f_1-f_3\delta c)A_{\mu}^2+(f_2-f_4\delta c)Tr(V_{\mu}^2)\right]\nonumber\\
\mathcal{L}^{(4)}&=&-\frac{1}{2}m^2(1-a\delta c)h^2+mh\left[(f_3-f_5\delta c)A_{\mu}^2+(f_4-f_6\delta c)Tr(V_{\mu}^2)\right]
+[g_0^i-(g_0^i)'\delta c]l_4^i\nonumber\\
\mathcal{L}^{(6)}&=&\frac{1}{2}(\partial_{\mu}h)^2-\frac{1}{6}am
h^3+\frac{1}{2}f_5h^2A_{\mu}^2+\frac{1}{2}f_6h^2Tr(V_{\mu}^2)\label{LEEWCL}\\
\mathcal{L}^{(8)}&=&-\frac{1}{12}bh^4+\frac{1}{6}m^{-1}h^3[f_7A_{\mu}^2+f_8Tr(V_{\mu}^2)]
+\frac{1}{2}(g_0^i)''h^2l_4^i+\frac{1}{2}g_2^km^{-2}(\partial_{\mu}h)(\partial_{\nu}h)l_2^{\mu\nu}\nonumber\\
\mathcal{L}^{(10)}&=&\frac{1}{2}(g_2^k)'m^{-3}(\partial_{\mu}h)(\partial_{\nu}h)hl_2^{k\mu\nu}\;.\nonumber
\end{eqnarray}
We will see that beyond order of $p^{10}$, it is impossible that
1-loop higgs corrections make contributions to $p^2$ and $p^4$ order
EWCL. When we are considering 1-loop corrections to $p^2$ and $p^4$
orders, all the coefficients of higher order operators take values
at tree level. In (\ref{LEEWCL}), there are 19 free parameters
undetermined, which are listed in Table~\ref{table-1} in more
detail.
\begin{table}[htb]
\begin{tabular}{|c|c|c|c|c|}
  \hline
  parameters &  & term & order & process \\ \hline
  m & higgs mass & $-\frac{1}{2}m^2h^2$ & 4 & higgs mass term \\
  $\mu$ & present energy scale & $\ln\frac{\mu^2}{m^2}$ &  &  \\
  $f_1$ & coupling & $Tr(TV_{\mu})Tr(TV^{\mu})$ & 2 & Z mass term  \\
  $f_2$ & coupling & $Tr(V_{\mu}V^{\mu})$ & 2 & $W^+W^-$ and Z mass term \\
  $f_3$ & coupling & $hTr(TV_{\mu})Tr(TV^{\mu})$ & 4 & $h\rightarrow ZZ$ \\
  $f_4$ & coupling & $hTr(V_{\mu}V^{\mu})$ & 4 & $h\rightarrow W^+W^-$ and ZZ \\
  $g_0^i$ & coupling & $l_4^i$ & 4 & EWCL \\
  $(g_0^i)'$ & coupling & $hl_4^i$ & 6 & EWCL \\
  $\frac{1}{2}f_5$ & coupling & $h^2Tr(TV_{\mu})Tr(TV^{\mu})$ & 6 & $hh\rightarrow ZZ$\\
  $\frac{1}{2}f_6$ & coupling & $h^2Tr(V_{\mu}V^{\mu})$ & 6 & $hh\rightarrow W^+W^-$ and ZZ \\
  $\frac{1}{6}a$ & coupling & $h^3$ & 6 & three higgs interaction \\
  $\frac{1}{6}f_7$ & coupling & $h^3Tr(TV_{\mu})Tr(TV^{\mu})$ & 8 & $hhh\rightarrow ZZ$ \\
  $\frac{1}{6}f_8$ & coupling & $h^3Tr(V_{\mu}V^{\mu})$ & 8   & $hhh\rightarrow W^+W^-$ and ZZ \\
  $\frac{1}{12}b$ & coupling & $h^4$ & 8 & four higgs interaction \\
  $g_2^1$ & coupling & $\partial_{\mu}h\partial_{\nu}hTr(TV^{\mu})Tr(TV^{\nu})$ & 8 & $hh\rightarrow ZZ$ \\
  $g_2^2$ & coupling & $\partial_{\mu}h\partial_{\nu}hTr(V^{\mu}V^{\nu})$ & 8 & $hh\rightarrow W^+W^-$ and ZZ \\
  $\frac{1}{2}(g_0^i)''$ & coupling & $h^2l_4^i$ & 8 & $hh\rightarrow EWCL$ \\
  $(g_2^1)'$ & coupling & $h\partial_{\mu}h\partial_{\nu}hTr(TV^{\mu})Tr(TV^{\nu})$ & 10 & $hhh\rightarrow ZZ$ \\
  $(g_2^2)'$ & coupling & $h\partial_{\mu}h\partial_{\nu}hTr(V^{\mu}V^{\nu})$ & 10 & $hhh\rightarrow W^+W^-$ and ZZ \\
  \hline
\end{tabular}
\vspace{0.2cm} \caption{The summary of parameters appeared in EEWCL}
\label{table-1}
\end{table}
\subsection{integrate out higgs up to 1-loop}
Different functional approaches of integrating out a scalar field
are used in Ref.~\cite{Zuk85,Chan85,Cheyette88}. To integrate out
higgs, we use the method of loop expansion in Ref.~\cite{Jackiw74},
but for our purpose now, the procedure will be slightly different.
In the traditional loop expansion  to integrate out some field
$\phi$, one first calculate the effect action loop by loop, then use
the solution of the stationery equation for this effective action to
subtract out $\phi$ field in the theory. Infinities originated from
loop integration will appear in the classic solution of the
stationery equation, which has to be renormalized. Technically these
infinities could be canceled by the corresponding counter terms in
the theory after $\phi$ is integrated out. However, this kind of
renormalization does not involve counter terms of the original
theory and thus loses information of loop corrections from couplings
denoting interactions between $\phi$ field and the rest part of the
theory, i.e., all the loop corrections are represented by these
couplings in the secondary theory after $\phi$ is integrated out.
Even we achieve to absorb all divergences from loops of $\phi$ field
into redefinitions of these secondary coefficients, the renormalized
coefficients are only related to bare ones in original theory with
$\phi$. To explicitly conserve contributions from the original
couplings of interactions involving $\phi$ field, both at tree level
and at loop level, we consider into theory the renormalization
effects of the couplings in original theory involving $\phi$ field
which should play their own roles in cancellation of divergences.
According to above analysis, we alter the procedure of integrating
out higgs by interchanging computation of searching the solution of
the stationery equation and of performing renormalization, i.e.
first we do 1-higgs-loop renormalization of EEWCL (\ref{EEWCL}),
then get the classic solution of the stationery equation of the
effective action, substitute it back into the Lagrangian, we
complete the 1-loop integration of higgs. In present procedure, all
divergences from loops can be absorbed into renormalized
coefficients of EEWCL, which represent physical
interactions among gauge bosons, goldstones and higgs. \\
In our computation system, we take low energy expansion which
assumes smaller coefficients of higher order operators. On the other
hand, these higher order operators contribute through loop
correction, which would receive an extra loop factor of $1/16\pi^2$,
hence makes the contribution from loops even smaller. This is the
reason we only take one loop approximation in the following of
calculations with higher loops effect omitted which will greatly
simplifies our computation work.

Effective action up to 1-loop is given by
\begin{eqnarray}
\Gamma^{1loop}&=&\int d^4x \mathcal{L}_{EEWCL}+\frac{i}{2}\ln
Det\hat D\;.
\end{eqnarray}
What remains to be done to include the 1-loop contribution in the
low energy expansion is to evaluate the determinant of the
differential operator $\hat D$, which is defined as
\begin{eqnarray}
\hat D(x,y)\equiv \frac{\delta^2 S}{\delta h(x)\delta h(y)}\;,
\end{eqnarray}
where $\hat D$ is given by,
\begin{eqnarray}
D=-(\partial^2+m^2-A+C^{\mu\nu}\partial_{\mu}\partial_{\nu})\;,
\end{eqnarray}
Where A and C are operators containing $p^2$ and $p^4$
contributions,
\begin{eqnarray}
A&=&-amh+f_5A_{\mu}^2+f_6Tr(V_{\mu}^2)-bh^2+f_7m^{-1}hA_{\mu}^2+f_8m^{-1}h Tr(V_{\mu}^2)+m^{-2}(g_0^i)''l_4^i\nonumber\\
C^{\mu\nu}&=&g_2^km^{-2}l^{k\mu\nu}_2+(g_2^k)'m^{-3}hl^{k\mu\nu}_2\;,
\end{eqnarray}

We take  dimensional regularization scheme to do the one loop
calculation, this choice can make us free of power type divergence
terms which can avoid the possible fault estimations \cite{London}
and their disturbance to our power countings. Then
\begin{eqnarray}
\frac{i}{2}\ln Det\hat D&=&\int d^4x\int \frac{\mu^{4-D}d^D
k}{(2\pi)^{D}}\langle x|k\rangle \ln det\hat D(\partial)\langle
k|x\rangle\nonumber\\
&=&\frac{i}{2}\int d^4x\int \frac{\mu^{4-D}d^D k}{(2\pi)^{D}}
[tr\ln\hat
D(\partial+ik)-\ln(k^2-m^2)]+Const\nonumber\\
&=&\frac{i}{2}\int d^4x\int \frac{\mu^{4-D}d^D k}{(2\pi)^{D}} \ln\left(1+\frac{-\partial^2-C^{\mu\nu}\partial_{\mu}\partial_{\nu}-2ik_{\mu}(C^{\mu\nu}+g^{\mu\nu})\partial_{\nu}+C^{\mu\nu}k_{\mu}k_{\nu}+A}{k^2-m^2}\right)\nonumber\\
&=&\frac{i}{2}\int d^4x\int \frac{\mu^{4-D}d^D
k}{(2\pi)^{D}}\left[\frac{A+C^{\mu\nu}k_{\mu}k_{\nu}}{k^2-m^2}-\frac{(A+C^{\mu\nu}k_{\mu}k_{\nu})^2}{2(k^2-m^2)^2}+O(p^6)\right]\nonumber\\
&=&\frac{1}{32\pi^2}\bigg[-(L+1)m^2A-\frac{(L+3/2)m^4}{4}C^{\mu}_{\mu}
+\frac{L}{2}A^2\nonumber\\
&&+\frac{(L+1)m^2}{2}AC^{\mu}_{\mu}+\frac{(L+3/2)m^4}{16}\left[(C^{\mu}_{\mu})^2+2(C^{\mu\nu})^2\right]+O(p^6)\bigg]
\label{lnD}\;,
\end{eqnarray}
In which $L$ is defined as£º
\begin{eqnarray}
L&\equiv&\frac{1}{\epsilon}-\gamma+\ln\frac{4\pi\mu^2}{m^2}
\end{eqnarray}
$\mu$ is energy scale appeared specially in dimensional
regularization, $A_2$, $A_4$ and $C_2$, $C_4$ denote $p^2$ and $p^4$
order parts in $A$ and $C$, respectively. In above calculation
following conventions are used,
\begin{eqnarray}
2\epsilon=4-D\hspace{0.8cm} \Gamma(z)=\int dt
e^{-t}t^{z-1}\hspace{0.8cm}\Gamma(z+1)=z\Gamma(z)\hspace{0.8cm}
\Gamma(\epsilon)&=&\frac{1}{\epsilon}-\gamma+O(\epsilon)\;.
\end{eqnarray}

1-loop renormalized result for $p^2$ and $p^4$ order Lagrangian
are listed below:
\begin{itemize}
\item $p^2$½×
\begin{eqnarray}
\mathcal{L}^{(2)}&=&m^2\left[\bar f_1A_{\mu}^2+\bar
f_2Tr(V_{\mu}^2)\right]\nonumber\\
&=&m^2\left[(f_1+\delta f_1)A_{\mu}^2+(f_2+\delta
f_2)Tr(V_{\mu}^2)\right]\label{p^2lag}\;,
\end{eqnarray}
\begin{eqnarray}
&&\hspace{-1cm}\bar
f_1=f_1+\frac{1}{32\pi^2}\left[-\frac{(L+3/2)}{4}g_2^1
-(L+1)(f_5+af_3)\right]\nonumber\\
&&\label{p^2correction}\\
&&\hspace{-1cm}\bar
f_2=f_2+\frac{1}{32\pi^2}\left[-\frac{(L+3/2)}{4}g_2^2 -(
L+1)(f_6+af_4)\right]\;,\nonumber
\end{eqnarray}
\item $p^4$½×
\begin{eqnarray}
\mathcal{L}^{(4)}&=&-\frac{1}{2}m^2_hh^2+m\bar f_3l_4^A+\bar
f_4l_4^v+\bar g_0^il_4^i\label{(p^4)lag}\\
&=&-\frac{1}{2}(m^2+\delta m^2)h^2+m(f_3+\delta f_3)l_4^A+
(f_4+\delta f_4)l_4^v+(g_0^i+\delta g_0^i)l_4^i\nonumber\;,
\end{eqnarray}
In which $l_4^i$, $i=1,2,\ldots,14$ denote the fourteen $p^4$ order
EWCL operators.
\begin{eqnarray}
m^2_h&=&m^2\left[1-\frac{1}{16\pi^2}(L+1)(a^2+b)-\frac{a^2}{32\pi^2}\right]\nonumber\\
\bar f_3&=&f_3+\frac{1}{32\pi^2}\left[-(L+1)f_7-\frac{L+3/2}{4}(g_2^1)'-\frac{L+1}{2}ag_2^1-(2L+1)af_5\right]\nonumber\\
\bar
f_4&=&f_4+\frac{1}{32\pi^2}\left[-(L+1)f_8-\frac{L+3/2}{4}(g_2^2)'-\frac{L+1}{2}ag_2^2-(2L+1)af_6\right]\nonumber\\
\label{correctionI}\,~
\end{eqnarray}
\begin{eqnarray}
\bar
g_0^4&=&g_0^4+\frac{1}{32\pi^2}\bigg[-(L+1)\left[(g_0^4)''+a(g_0^4)'\right]+\frac{L+3/2}{8}(g_2^2)^2\bigg]\nonumber\\
\bar
g_0^6&=&g_0^6+\frac{1}{32\pi^2}\bigg[-(L+1)\left[(g_0^6)''+a(g_0^6)'\right]+\frac{L+3/2}{4}g_2^1g_2^2\bigg]\nonumber\\
\bar
g_0^5&=&g_0^5+\frac{1}{32\pi^2}\bigg[-(L+1)\left[(g_0^5)''+a(g_0^5)'\right]+\frac{L}{2}(f_6)^2+\frac{L+1}{2}f_6g_2^2
+\frac{L+3/2}{16}(g_2^2)^2\bigg]\nonumber\\
\bar
g_0^7&=&g_0^7+\frac{1}{32\pi^2}\bigg[-(L+1)\left[(g_0^7)''+a(g_0^7)'\right]+Lf_5f_6+\frac{L+1}{2}(f_5g_2^2+f_6g_2^1)
+\frac{L+3/2}{8}g_2^1g_2^2\bigg]\nonumber\\\nonumber\\
\bar
g_0^{10}&=&g_0^{10}+\frac{1}{32\pi^2}\bigg[-(L+1)\left[(g_0^{10})''+a(g_0^{10})'\right]+\frac{L}{2}(f_5)^2+\frac{L+1}{2}f_5g_2^1
+\frac{3(L+3/2)}{16}(g_2^1)^2\bigg]\nonumber\\
\label{correction2}
\end{eqnarray}
\begin{eqnarray}
\bar g_0^i&=&g_0^i+\frac{1}{32\pi^2}\bigg[-(L+1)\left[(g_0^4)''+a(g_0^i)'\right]\bigg]\label{correction3}\\
i&=&1,2,3,8,9,11,12,13,14\nonumber
\end{eqnarray}
\end{itemize}
All those $\delta f_i=\bar f_i-f_i$ and $\delta g_0=\bar g_0-g_0$
terms denote 1-loop level contributions to $p^4$ order EWCL
couplings from higher order interactions among higgs, goldstones
and gauge bosons, while $p^4$ operators in (\ref{(p^4)lag}) can
only contribute to $p^4$ order EWCL couplings at tree level. From
(\ref{(p^4)lag}), we get the stationery equation for $h_c$,
\begin{eqnarray}
h_c=\frac{m}{m_h^2}[{\bar f_3}Tr(TV_{\mu})Tr(T^{\mu})+{\bar
f_4}Tr(V_{\mu}V^{\mu})]\;.
\end{eqnarray}
Note this solution is expressed by the 1-higgs-loop renormalized
coefficients of the theory before higgs is integrated out.
Substitute $h_c$ back into (\ref{(p^4)lag}), we get the tree level
contribution from $p^4$ order higgs interacting with gauge bosons,
\begin{eqnarray}
\mathcal{L}^{(4)}_{EWCL}&=&\frac{m^2}{2m_h^2}\left[(\bar
f_3)^2l_4^{10}+2\bar f_3\bar f_4l_4^{7}+(\bar
f_4)^2l_4^{5}\right]\label{treecontribution}\;.
\end{eqnarray}

Hence, after higgs is integrated out at 1-loop level in the low
energy expansion up to $p^4$, we find out following final EWCL
which include contributions from higgs interactions,
\begin{eqnarray}
\mathcal{L}_{EWCL}&=&m^2\left[\bar f_1Tr(TV_{\mu})Tr(TV^{\mu})+\bar
f_2Tr(V_{\mu}^2)\right]+(g_0^i+\Delta
g_0^i)l_4^i\label{corrections}\;,
\end{eqnarray}
with $\Delta g_0^i$ denoting the contribution from integrating out
higgs,
\begin{eqnarray}
\Delta g_0^5&=&\frac{1}{2}f_4^2+f_4\delta
f_4-\frac{1}{2}(f_4)^2\frac{\delta m^2}{m^2}+\delta g_0^5\nonumber\\
\Delta g_0^7&=&f_3f_4+f_3\delta f_4+f_4\delta f_3-f_3f_4\frac{\delta
m^2}{m^2}+\delta g_0^7\nonumber\\
\Delta g_0^5&=&\frac{1}{2}f_3^2+f_3\delta
f_3-\frac{1}{2}(f_3)^2\frac{\delta m^2}{m^2}+\delta
g_0^{10}\nonumber\\
\Delta g_0^j&=&\delta g_0^j, j\neq 5,7,10\label{Delta}\;.
\end{eqnarray}
In following discussion we will use the redefined $\tilde g_0^i$
containing full contribution from integrating out higgs up to one
loop level,
\begin{eqnarray}
\tilde g_0^i\equiv g_0^i+\Delta g_0^i
\end{eqnarray}
In (\ref{Delta}), terms with $\delta$ represent one loop
contribution, which should be small, since they appear as a product
of the loop factor $1/16\pi^2$ and higher order coefficients. Since
we focus on the effects of higgs, the loop calculation does not
include radiative corrections from other particles existed in the
theory, the gauge bosons and goldstone bosones are totally viewed as
a classic external source, i.e., there's no consideration about
internal line or loops of gauge bosons and goldstone bosons. Loop
effects from these particles can be viewed as backgrounds when we
compare how higgs and some other new particle interactions will
alter EWCL couplings through loops. Those contributions of other
possible new particles will be investigated in separated papers.
\section{Discussions and Summary}

Before starting the detail discussions, we first make some general
analysis. First the coefficients in EWCL have their bare values
$g_0^i$ and corrections $\Delta g_0^i$ from higgs. The bare part
$g_0^i$ is given in original EEWCL (\ref{LEEWCL}) and is
independent of physics related to higgs. Since this paper focus on
the higgs contributions part, we need to invent some arguments
judging the smallness for values of those bare coefficients. We
take assumption that higgs in reality will {\it really be} the
next new particle we find in future experiment, then we could
expect that it must plays an important role in the physics just
below its threshold which implies at this scale some effects from
bare coefficients are small compare to those from higgs. It is
under this assumption, the contribution to some coefficients from
higgs should be larger than its bare part: $\Delta g_0^i\gg g_0^i$
for some $i$ and then we can ignore the corresponding bare part
contributions.

From (\ref{correctionI}), (\ref{correction2}), (\ref{correction3}),
(\ref{treecontribution}) and (\ref{corrections}), we see higgs
contribution to EWCL coefficients starts from $p^4$ operators at
tree level, and higher order operators contribute in two ways:
1.induce direct corrections through higgs loop, 2. induce correction
to $p^4$ Lagrangian (\ref{(p^4)lag}). If we omit all the loop
corrections, we see higgs contribution concentrates in three terms:
$\mathcal{L}_5$,$\mathcal{L}_7$ and $\mathcal{L}_{10}$, from
$h\rightarrow ZZ$ and $h\rightarrow W^{\pm}$ decay. Thus, in tree
level estimation, these two channels are important in phenomenology
to find out the connection between higgs and EWCL couplings. A
similar conclusion is presented in Ref.~\cite{zhangbin03}. The most
important loop correction comes from those $p^6$ operators with
couplings $f_5$, $f_6$ and $a$, the first two couplings include
$hh\rightarrow ZZ$ and $hh\rightarrow W^{\pm}$ scattering process,
$a$ is the coupling of three higgs self-interaction. From
(\ref{(p^4)lag}) and (\ref{correctionI}), we see $f_5$, $f_6$ and
$a$ all contribute through loop correction to $p^4$ operators
$h[Tr(TV_{\mu})]^2$ and $hTr(V_{\mu}V^{\mu})$, which contribute to
$\mathcal{L}_5$,$\mathcal{L}_{7}$ and $\mathcal{L}_{10}$ at tree
level; from (\ref{correction2}) we see $f_5$, $f_6$ contribute
directly to $\mathcal{L}_5$,$\mathcal{L}_{7}$ and
$\mathcal{L}_{10}$. Hence these three EWCL couplings should be most
important in phenomenology when testing higgs signals connected to
EWCL. Further, (\ref{p^2correction}) tells us that $p^2$ EWCL
operators in (\ref{p^2lag}) are altered through loop correction by
$g_2^i$, $f_5\!+\!af_3$ and $f_6\!+\!af_4$, involving
$h,hh\rightarrow VV$ and three higgs self interaction process. Since
$f_6$ gives correction to $f^2$ in
$-\frac{f^2}{4}Tr(V_{\mu}V^{\mu})$, $f_5$ gives correction to
$\beta_1f^2$ in $\frac{\beta_1f^2}{4}[Tr(TV_{\mu})]^2$, combine them
together we can fix $\beta_1$ which is related to T parameter given
by M.E.Peskin and T.Takeuchi \cite{peskin92} through relation
$\alpha T=2\beta_1$ \cite{Appelquist93}, $f_5$ and $f_6$ might alter
the value of T through higgs loop correction.
\subsection{higgs decay and four-gauge-boson coupling}
$f_3$ and $f_4$ are related to couplings $g_{hWW}$ and $g_{hZZ}$
in (\ref{LEEWCL}). Take unitary gauge, they are given by,
\begin{eqnarray}
g_{hWW}&=&-\frac{e^2}{s^2}f_4\nonumber\\
g_{hZZ}&=&-\frac{e^2}{s^2c^2}(f_3+\frac{1}{2}f_4)
\end{eqnarray}
And the partial decay width of higgs is given by
\begin{eqnarray}
\Gamma_{h\rightarrow
ZZ}&=&\frac{(2f_3+f_4)^2 e^4m_h}{32s^4c^4}(1-\frac{4m^2_Z}{m_h^2})^{\frac{1}{2}}\nonumber\\
\Gamma_{h\rightarrow WW}&=&\frac{(f_4)^2
e^4m_h}{32s^4}(1-\frac{4m^2_W}{m_h^2})^{\frac{1}{2}}\;,
\end{eqnarray}
or
\begin{eqnarray}
(f_4)^2&=&\frac{32s^4}{e^4m_h}(1-\frac{4m_W^2}{m_h^2})^{-\frac{1}{2}}\Gamma_{h\rightarrow
WW}\nonumber\\
f_3f_4&=&\frac{8s^4}{e^4m_h}\bigg[c^4(1-\frac{4m_Z^2}{m_h^2})^{-\frac{1}{2}}\Gamma_{h\rightarrow
ZZ}-(1-\frac{4m_W^2}{m_h^2})^{-\frac{1}{2}}\Gamma_{h\rightarrow
WW}\bigg]\nonumber\\
(f_3)^2&=&\frac{2s^4}{e^4m_h}\bigg[(1-\frac{4m_W^2}{m_h^2})^{-\frac{1}{2}}\Gamma_{h\rightarrow
WW}-2c^4(1-\frac{4m_Z^2}{m_h^2})^{-\frac{1}{2}}\Gamma_{h\rightarrow
ZZ}\nonumber\\
&&+c^8(1-\frac{4m_W^2}{m_h^2})^{\frac{1}{2}}(1-\frac{4m_Z^2}{m_h^2})^{-1}\frac{(\Gamma_{h\rightarrow
ZZ})^2}{\Gamma_{h\rightarrow WW}}\bigg]\label{treeandwidth}\;,
\end{eqnarray}\\
where $s=\sin\theta_w$ and $c=\cos\theta_w$. Approximation $f_3\ll
f_4$ is used, due to the fact that $f_3$ induces explicit
custodial symmetry breaking, which is small in
reality\cite{longhitano80}. Substitute this result into
(\ref{treecontribution}), we get tree level estimation for the
contribution from higgs decaying into VV to four-gauge-boson
coupling, i.e., $l_4^{5}$, $l_4^{7}$ and $l_4^{10}$
\begin{eqnarray}
\mathcal{L}_{EWCL}\bigg|_{5,7,10}&=&\frac{32s^4}{e^4m_h}\bigg[(1-\frac{4m_W^2}{m_h^2})^{-\frac{1}{2}}\Gamma_{h\rightarrow
WW}-2c^4(1-\frac{4m_Z^2}{m_h^2})^{-\frac{1}{2}}\Gamma_{h\rightarrow
ZZ}\nonumber\\
&&+c^8(1-\frac{4m_W^2}{m_h^2})^{\frac{1}{2}}(1-\frac{4m_Z^2}{m_h^2})^{-1}\frac{(\Gamma_{h\rightarrow
ZZ})^2}{\Gamma_{h\rightarrow
WW}}\bigg][Tr(TV_{\mu})Tr(TV_{\nu})]^2\nonumber\\
&&+\frac{16s^4}{e^4m_h}\bigg[c^4(1-\frac{4m_Z^2}{m_h^2})^{-\frac{1}{2}}\Gamma_{h\rightarrow
ZZ}-(1-\frac{4m_W^2}{m_h^2})^{-\frac{1}{2}}\Gamma_{h\rightarrow
WW}\bigg]Tr(V_{\mu}V^{\mu})[Tr(TV_{\nu})]^2\nonumber\\
&&+\frac{4s^4}{e^4m_h}(1-\frac{4m_W^2}{m_h^2})^{-\frac{1}{2}}\Gamma_{h\rightarrow
WW}[Tr(V_{\mu}V^{\mu})]^2\label{width}\;.
\end{eqnarray}
Thus, the coefficient of $l_4^5\equiv [Tr(V_{\mu}V^{\mu})]^2$ is
most sensitive to the partial decay width of $h\rightarrow WW$. In
the energy scale below higgs mass, (\ref{width}) can be used to
estimate the value of the coupling $g_{hVV}$ or the partial width
of $h\rightarrow VV$, once four-gauge-boson coupling is obtained
in the lab.
\subsection{higgs mass dependence of 1-loop correction}
Higgs mass dependence of each dimension zero coefficient in
(\ref{corrections}) is determined by $ L$ with
\begin{eqnarray}
m^2\frac{d L}{dm^2}=-1
\end{eqnarray}
The result is given in Table \ref{table-2}. We can see only $\bar
f_1$ and $\bar f_2$ accept $p^6$ order corrections. One loop
correction for other couplings starts from $p^8$.
\begin{table}[htb]
\begin{tabular}{|c|c|c|c|c|}
  \hline
  $C$ &  $\mbox{$16\pi^2$}\frac{dC}{d\ln m}|_{p^6}$&$\mbox{$16\pi^2$}\frac{dC}{d\ln m}|_{p^8}$&
  $\mbox{$16\pi^2$}\frac{dC}{d\ln m}|_{p^{10}}$&$\mbox{$16\pi^2$}\frac{dC}{d\ln m}|_{p^{12}}$\\ \hline\hline
  $\bar f_1$ & $f_5$&$\frac{g_2^1}{4}$&$af_3$&\\
  $\bar f_2$ & $f_6$&$\frac{g_2^2}{4}$&$af_4$&\\
  $\bar f_3$ &  &$f_7$&$\frac{(g_2^1)'}{4}$&$2af_5$\\
  $\bar f_4$ &  &$f_8$&$\frac{(g_2^2)'}{4}$&$2af_6$\\
  $\tilde g_0^4$ & &$(g_0^4)''$&&$a(g_0^4)'$ \\
  $\tilde g_0^6$ &  &$(g_0^6)''$&&$a(g_0^4)'$\\
  $\tilde g_0^5$ & &$(g_0^5)''$&&$f_4f_8+a(g_0^5)'-\frac{(f_6)^2}{2}$\\
  $\tilde g_0^7$ & &$(g_0^7)''$&&$f_3f_7+f_4f_8+a(g_0^7)'+f_5f_6$\\
  $\tilde g_0^{10}$ & &$(g_0^{10})''$&&$f_3f_7+a(g_0^{10})'-\frac{(f_5)^2}{2}$\\
  $\tilde g_0^i$ &  &$(g_0^i)''$&&$a(g_0^i)'$\\
  \hline
  $C$ &  $\mbox{$16\pi^2$}\frac{dC}{d\ln m}|_{p^{14}}$&$\mbox{$16\pi^2$}\frac{dC}{d\ln m}|_{p^{16}}$&
  $\mbox{$16\pi^2$}\frac{dC}{d\ln m}|_{p^{18}}$&$\mbox{$16\pi^2$}\frac{dC}{d\ln m}|_{p^{20}}$\\\hline\hline
  $\bar f_1$ &&&&\\
  $\bar f_2$ &&&&\\
  $\bar f_3$ &$\frac{ag_2^1}{2}$&&&\\
  $\bar f_4$ &$\frac{ag_2^2}{2}$&&&\\
  $\tilde g_0^4$ &&$-\frac{(g_2^2)^2}{8}$&& \\
  $\tilde g_0^6$ &&$-\frac{g_2^1g_2^2}{4}$&&\\
  $\tilde g_0^5$ &$\frac{f_6g_2^2}{2}-\frac{f_4(g_2^2)'}{4}$&$2af_4f_6+\frac{b(f_4)^2}{2}-\frac{(g_2^2)^2}{16}$&$\frac{af_4g_2^2}{2}$&$-\frac{(af_4)^2}{2}$ \\
  $\tilde g_0^7$ &$\frac{(f_5g_2^1+f_6g_2^2)}{2}-\frac{[f_3(g_2^2)'+f_4(g_2^1)']}{4}$&$2a(f_3f_6+f_4f_5)+bf_3f_4-\frac{g_2^1g_2^2}{8}$&$\frac{a(f_3g_2^2+f_4g_2^1)}{2}$&$-af_3f_4$ \\
  $\tilde g_0^{10}$ &$\frac{f_5g_2^1}{2}-\frac{f_3(g_2^1)'}{4}$&$2af_3f_5+\frac{b(f_3)^2}{2}-\frac{3(g_2^1)^2}{16}$&$\frac{af_3g_2^1}{2}$&$-\frac{(af_3)^2}{2}$ \\
  $\tilde g_0^i$&&&&\\
  \hline
\end{tabular}
\vspace{0.2cm} \caption{The summary of higgs mass dependence of
coefficients appeared in (\ref{corrections})} \label{table-2}
\end{table}

Now we turn to the higgs mass dependence of certain terms related to
T and S parameters in (\ref{corrections}).
\begin{eqnarray}
\frac{d\bar f_1}{dm^2}=\frac{1}{32\pi^2}\frac{1}{m^2}\left[\frac{1}{4}g^1_2+f_5+af_3\right]\\
\frac{d\bar
f_2}{dm^2}=\frac{1}{32\pi^2}\frac{1}{m^2}\left[\frac{1}{4}g^2_2+f_6+af_4\right]\;.
\end{eqnarray}
Because $af_3$,$g_2^1$ and $af_4$,$g_2^2$ are respectively higher in
order than $f_5$ and $f_6$, it's reasonable to suppose $g_2^1,
af_3\ll f_5$, $(g_2^2), af_4\ll f_6$, thus they are omitted. Since
$-\beta_1=\bar f_1/\bar f_2$,
\begin{eqnarray}
\alpha m^2\frac{dT}{dm^2}=2\frac{d\beta_1}{dm^2}=\frac{2}{\bar
f_2^2}(\bar f_2\frac{d\bar f_1}{dm^2}-\frac{\bar f_1d\bar
f_2}{dm^2})=\frac{1}{16\pi^2}\frac{1}{\bar f_2}[f_5+\beta_1f_6]\;.
\end{eqnarray}
For a given value of higgs mass, since $\bar f_2\equiv
-f^2/4<0$\cite{Appelquist93}, when
\begin{eqnarray}
f_5+\beta_1f_6<0\label{f_3f_4}\;.
\end{eqnarray}
T parameter increases when higgs mass increases, or present energy
scale decreases far away from higgs mass, which means a heavier
higgs allows larger value of T parameter, which is consistent with
SM data in Ref.~\cite{zhangbin03,Kuang03}. S parameter is related to
$g_0^1$, which has mass dependence as
\begin{eqnarray}
m^2\frac{dS}{dm^2}=m^2\frac{dg_0^1}{dm^2}=\frac{1}{32\pi^2}[(g_0^1)''+a(g_0^1)']\;.
\end{eqnarray}
SM data in Ref.~\cite{zhangbin03,Kuang03} tells smaller value of S
is consistent with the existence of heavier higgs for better data
fitting result. Thus $(g_0^1)''+a(g_0^1)'<0$.

\subsection{higgs mass limit}
For a given value of $\mu$, here we test a assumption that when
higgs mass goes to infinity, or when $\mu\ll m$, all the
contributions from higgs loop to EWCL couplings should decrease
when higgs mass increases. This is a reasonable assumption because
the inference from higgs should become smaller and smaller when it
is farer and farer away from present energy scale. That is to say
the coefficients should have such mass dependence
\begin{eqnarray}
\frac{d\delta C}{dm^2}\delta C<0\;.
\end{eqnarray}
However, this condition can only be satisfied when there is a factor
of minus power of mass. From above one loop result we see there is
no such term which accepts this restriction. This is due to the fact
that the only dimensional constant in EEWCL is higgs mass m (the
vacuum condensation v should be proportional to m), since classic
solution of higgs field is expanded as
\begin{eqnarray}
\tilde h\sim m^{-1}\mathcal{L}^{(2)}+m^{-3}\mathcal{L}^{(4)}+\ldots
\end{eqnarray}
Suppose a general form of EEWCL is written as
\begin{eqnarray}
\mathcal{L}_{EEWCL}=\sum\limits_nC_nm_h^x\tilde
h^y\mathcal{L}^{2z}\label{sample}
\end{eqnarray}
with relation $x+y+2z=4$. $\mathcal{L}$ denotes the external source
with momentum order $p^{2z}$. We know (\ref{sample}) can be expanded
in momentum order as
\begin{eqnarray}
\sum\limits_nC_n(p^{2y+2z}+p^{4y+2z}+\ldots)\label{expand2}
\end{eqnarray}

On the other hand, if we view it as a expansion of different
operators in $m^{-1}$,
\begin{eqnarray}
\sum\limits_nC_n\left[(\frac{1}{m})^{2y+2z-4}+(\frac{1}{m})^{4y+2z-4}+\ldots\right]\label{expand1}
\end{eqnarray}
Up to $p^4$, we have the relation
\begin{eqnarray}
O(m^{-1})=O(p)-4\leq 0
\end{eqnarray}
This means up to $p^4$ order there include only non-decoupling
effect of heavy higgs, while decoupling effect in the higgs mass
limit $m_h\rightarrow \infty$ would not show up until in $p^6$ and
higher orders.
\subsection{standard model higgs}
Although the title of this paper is called nonstandard higgs in
EWCL, since we have written down the most general form of EEWCL,
conventional standard model higgs is certainly included in our
theory. If we set as TABLE \ref{table-3}
\begin{table}[htb]\begin{center}
\begin{tabular}{c|c|c|c|c|c|c}
  \hline
  C &  $\bar f_2$ & $\bar f_4$ & $f_6$ & a & b& others\\\hline
  value & $-\tfrac{1}{32\lambda}$ & $-\tfrac{\sqrt{2}}{8}\tfrac{1}{\sqrt{\lambda}}$&$-\tfrac{1}{2} $& $6\sqrt{2\lambda}$ & $12\lambda$& 0\\\hline
\end{tabular}\end{center}
\vspace{0.2cm} \caption{The summary of coefficients appeared in
SM.}\label{table-3}
\end{table}
and all other couplings become zero, we go back to SM higgs
 in (\ref{SM}).  Then we can use result in previous subsections to get the result of integrating out SM higgs, which is
 listed in TABLE \ref{table-4}.
\begin{table}[htb]\begin{center}
\begin{tabular}{c|c|c|c|c}
  \hline
  C &  $~\bar f_2~$ & $~\bar f_4~$ & $~\tilde g_0^5~$ & others\\\hline
  $8\pi^2\tfrac{dC}{d\ln m}$ & $-1$ & $\tfrac{3}{2}$& $\tfrac{7}{32}$ & $0$\\\hline
\end{tabular}\end{center}
\vspace{0.2cm} \caption{The summary of higgs mass dependence of
coefficients appeared in SM.}\label{table-4}
\end{table}
Obviously higgs in SM cannot be decoupled from the rest of theory
even though it is very heavy. We see T parameter decreases when
higgs mass increases in SM according to condition (\ref{f_3f_4}).
And from Table~\ref{table-3} we see again the corrections have no
dependence on $\lambda$.

In Summary, we have investigated the possible effects from single
higgs to all $p^2$ and $p^4$ order coefficients in bosonic part of
EWCL. We have included in all possible higgs couplings which may
contribute to $p^2$ and $p^4$ order EWCL coefficients within one
higgs-loop precision. We find three terms, $\mathcal{L}_5$,
$\mathcal{L}_7$, $\mathcal{L}_{10}$ in EWCL are important, for
which the contributions from higgs can be further expressed in
terms of higgs partial decay width $\Gamma_{h\rightarrow ZZ}$ and
$\Gamma_{h\rightarrow WW}$. Higg mass dependence of the
coefficients are discussed and SM is one of special case in our
discussion.

\section*{Acknowledgments}
We specially thank Professor Hong-Jian He for helpful suggestions in
several useful references. This work was supported by National
Science Foundation of China (NSFC) and Specialized Research Fund for
the Doctoral Program of High Education of China.

\end{document}